\documentstyle[twoside]{article}

\catcode`\@=11
\long\def\@makefntext#1{
\protect\noindent \hbox to 3.2pt {\hskip-.9pt  
$^{{\eightrm\@thefnmark}}$\hfil}#1\hfill}		

\def\@makefnmark{\hbox to 0pt{$^{\@thefnmark}$\hss}}	
	
\def\ps@myheadings{\let\@mkboth\@gobbletwo
\def\@oddhead{\hbox{}
\rightmark\hfil\eightrm\thepage}   
\def\@oddfoot{}\def\@evenhead{\eightrm\thepage\hfil
\leftmark\hbox{}}\def\@evenfoot{}
\def\sectionmark##1{}\def\subsectionmark##1{}}

\oddsidemargin=\evensidemargin
\newcounter{sectionc}\newcounter{subsectionc}\newcounter{subsubsectionc}
\renewcommand{\section}[1] {\vspace{12pt}\addtocounter{sectionc}{1} 
\setcounter{subsectionc}{0}\setcounter{subsubsectionc}{0}\noindent 
	{\tenbf\thesectionc. #1}\par\vspace{5pt}}
\renewcommand{\subsection}[1] {\vspace{12pt}\addtocounter{subsectionc}{1} 
	\setcounter{subsubsectionc}{0}\noindent 
	{\bf\thesectionc.\thesubsectionc. {\kern1pt \bfit #1}}\par\vspace{5pt}}
\renewcommand{\subsubsection}[1] {\vspace{12pt}\addtocounter{subsubsectionc}{1}
	\noindent{\tenrm\thesectionc.\thesubsectionc.\thesubsubsectionc.
	{\kern1pt \tenit #1}}\par\vspace{5pt}}
\newcommand{\nonumsection}[1] {\vspace{12pt}\noindent{\tenbf #1}
	\par\vspace{5pt}}

\newcounter{appendixc}
\newcounter{subappendixc}[appendixc]
\newcounter{subsubappendixc}[subappendixc]
\renewcommand{\thesubappendixc}{\Alph{appendixc}.\arabic{subappendixc}}
\renewcommand{\thesubsubappendixc}
	{\Alph{appendixc}.\arabic{subappendixc}.\arabic{subsubappendixc}}

\renewcommand{\appendix}[1] {\vspace{12pt}
        \refstepcounter{appendixc}
        \setcounter{figure}{0}
        \setcounter{table}{0}
        \setcounter{lemma}{0}
        \setcounter{theorem}{0}
        \setcounter{corollary}{0}
        \setcounter{definition}{0}
        \setcounter{equation}{0}
        \renewcommand{\thefigure}{\Alph{appendixc}.\arabic{figure}}
        \renewcommand{\thetable}{\Alph{appendixc}.\arabic{table}}
        \renewcommand{\theappendixc}{\Alph{appendixc}}
        \renewcommand{\thelemma}{\Alph{appendixc}.\arabic{lemma}}
        \renewcommand{\thetheorem}{\Alph{appendixc}.\arabic{theorem}}
        \renewcommand{\thedefinition}{\Alph{appendixc}.\arabic{definition}}
        \renewcommand{\thecorollary}{\Alph{appendixc}.\arabic{corollary}}
        \renewcommand{\theequation}{\Alph{appendixc}.\arabic{equation}}
        \noindent{\tenbf Appendix \theappendixc #1}\par\vspace{5pt}}
\newcommand{\subappendix}[1] {\vspace{12pt}
        \refstepcounter{subappendixc}
        \noindent{\bf Appendix \thesubappendixc. {\kern1pt \bfit #1}}
	\par\vspace{5pt}}
\newcommand{\subsubappendix}[1] {\vspace{12pt}
        \refstepcounter{subsubappendixc}
        \noindent{\rm Appendix \thesubsubappendixc. {\kern1pt \tenit #1}}
	\par\vspace{5pt}}

\newcommand{\textlineskip}{\baselineskip=13pt}
\newcommand{\smalllineskip}{\baselineskip=10pt}

\def\eightcirc{
\begin{picture}(0,0)
\put(4.4,1.8){\circle{6.5}}
\end{picture}}
\def\eightcopyright{\eightcirc\kern2.7pt\hbox{\eightrm c}}

\def\abstracts#1#2#3{{
	\centering{\begin{minipage}{4.5in}\baselineskip=10pt\footnotesize
	\parindent=0pt #1\par 
	\parindent=15pt #2\par
	\parindent=15pt #3
	\end{minipage}}\par}} 

\renewenvironment{thebibliography}[1]
	{\frenchspacing
	 \ninerm\baselineskip=11pt
	 \begin{list}{\arabic{enumi}.}
	{\usecounter{enumi}\setlength{\parsep}{0pt}
	 \setlength{\leftmargin 12.7pt}{\rightmargin 0pt} 
	 \setlength{\itemsep}{0pt} \settowidth
	{\labelwidth}{#1.}\sloppy}}{\end{list}}

\newcounter{itemlistc}
\newcounter{romanlistc}
\newcounter{alphlistc}
\newcounter{arabiclistc}

\newcommand{\fcaption}[1]{
        \refstepcounter{figure}
        \setbox\@tempboxa = \hbox{\footnotesize Fig.~\thefigure. #1}
        \ifdim \wd\@tempboxa > 5in
           {\begin{center}
        \parbox{5in}{\footnotesize\smalllineskip Fig.~\thefigure. #1}
            \end{center}}
        \else
             {\begin{center}
             {\footnotesize Fig.~\thefigure. #1}
              \end{center}}
        \fi}

\newcommand{\tcaption}[1]{
        \refstepcounter{table}
        \setbox\@tempboxa = \hbox{\footnotesize Table~\thetable. #1}
        \ifdim \wd\@tempboxa > 5in
           {\begin{center}
        \parbox{5in}{\footnotesize\smalllineskip Table~\thetable. #1}
            \end{center}}
        \else
             {\begin{center}
             {\footnotesize Table~\thetable. #1}
              \end{center}}
        \fi}

\def\@citex[#1]#2{\if@filesw\immediate\write\@auxout
	{\string\citation{#2}}\fi
\def\@citea{}\@cite{\@for\@citeb:=#2\do
	{\@citea\def\@citea{,}\@ifundefined
	{b@\@citeb}{{\bf ?}\@warning
	{Citation `\@citeb' on page \thepage \space undefined}}
	{\csname b@\@citeb\endcsname}}}{#1}}

\newif\if@cghi
\def\cite{\@cghitrue\@ifnextchar [{\@tempswatrue
	\@citex}{\@tempswafalse\@citex[]}}
\def\citelow{\@cghifalse\@ifnextchar [{\@tempswatrue
	\@citex}{\@tempswafalse\@citex[]}}
\def\@cite#1#2{{$\null^{#1}$\if@tempswa\typeout
	{IJCGA warning: optional citation argument 
	ignored: `#2'} \fi}}

\def\pmb#1{\setbox0=\hbox{#1}
	\kern-.025em\copy0\kern-\wd0
	\kern.05em\copy0\kern-\wd0
	\kern-.025em\raise.0433em\box0}

\def\fnt#1#2{\footnotetext{\kern-.3em
	{$^{\mbox{\scriptsize #1}}$}{#2}}}

\font\tenrm=cmr10
\font\tenit=cmti10 
\font\tenbf=cmbx10
\font\bfit=cmbxti10 at 10pt
\font\ninerm=cmr9

\font\eightrm=cmr8

\def\qed{\hbox{${\vcenter{\vbox{			
   \hrule height 0.4pt\hbox{\vrule width 0.4pt height 6pt
   \kern5pt\vrule width 0.4pt}\hrule height 0.4pt}}}$}}

\begin{document}

\begin{flushright}                                
UMN-D-00-6 \\ October 2000 \\ \vspace{0.3in}
\end{flushright}


\normalsize\textlineskip
\setcounter{page}{1}



\centerline{\bf APPLICATION OF DISCRETE LIGHT-CONE QUANTIZATION}
\vspace*{0.035truein}
\centerline{\bf TO YUKAWA THEORY IN FOUR DIMENSIONS%
\footnote{To appear in the proceedings of               
DPF2000, Columbus, Ohio, August 9-12, 2000.}%
}

\vspace*{0.37truein}
\centerline{\footnotesize J.R. HILLER%
\footnote{\baselineskip=14pt                            
Work supported in part by the Department of Energy,
contract DE-FG02-98ER41087.}%
}
\vspace*{0.015truein}
\centerline{\footnotesize\it Department of Physics, University of Minnesota-Duluth}
\baselineskip=10pt
\centerline{\footnotesize\it Duluth, Minnesota~~55812, USA}

\vspace*{0.21truein}
\abstracts{The numerical technique of discrete light-cone 
quantization (DLCQ) is applied to a single-fermion truncation 
of Yukawa theory in four dimensions.  The truncated theory is 
regulated by three Pauli--Villars bosons, which are introduced 
directly in the DLCQ Fock-state basis.  A special form of the 
Lanczos diagonalization algorithm is used to handle the indefinite 
metric.  Renormalization is done nonperturbatively.}{}{}

\vspace*{1pt}\textlineskip
\section{Introduction}
\noindent
Methods for the nonperturbative numerical solution of 
light-cone-quantized quantum field theories have progressed to the 
point where they are applicable in four dimensions.  The use of
light-cone coordinates\cite{Dirac} makes possible a meaningful
Fock-state expansion, in which no disconnected vacuum contributions
appear.  The Fock-state wave functions in the expansions are
obtained by solving a Hamiltonian eigenvalue problem.\cite{DLCQreview}
A standard technique for solving such a problem is discrete
light-cone quantization (DLCQ).\cite{PauliBrodsky,DLCQreview}
The wave functions are evaluated at discrete momentum values
$p^+\equiv E+p_z = n\pi/L$, 
$\vec{p}_\perp\equiv (p_x,p_y) = \vec{n}_\perp \pi/L_\perp$,
with $n$, $n_x$, and $n_y$ integers and $L$ and $L_\perp$
length scales.  
The coupled integral equations that comprise the eigenvalue
problem become a matrix diagonalization problem where
trapezoidal sums approximate integrals.

To properly formulate such a matrix problem as an approximation
to a four-dimensional field theory, one must include a regularization
scheme and perhaps additional cutoffs that yield a finite
matrix problem.  One must also include a renormalization scheme
to determine bare parameters.  These steps have been carried
out for simple models\cite{PV1,PV2} and are now being applied
to Yukawa theory.\cite{Adelaide}  The regularization scheme is
based on the introduction of Pauli--Villars particles,\cite{PauliVillars}
including some with negative norm, and of a simple mass counterterm.
The theory is then finite before discretization.  A cutoff
on the light-cone energy $(m^2+p^2_\perp)/p^+$ is
used to limit the transverse momentum range and produce a matrix
representation of finite size.  
Renormalization is done nonperturbatively, by fixing computable
quantities to ``data.''

The original theory is recovered in the following sequence of
limits.  First, the numerical limit of infinite longitudinal
and transverse resolutions is taken at
fixed cutoff and fixed Pauli--Villars masses.  Next,
the cutoff is removed, and, finally, the Pauli--Villars
masses are taken to infinity.  Calculations done to date\cite{PV1,PV2}
do not show a great sensitivity to numerical resolution, above
a modest threshold, so that the basis sizes used in the
matrix problem have been manageable.  The largest basis
used is approximately 10.5 million states.

\section{Yukawa Theory}
\noindent
The light-cone Hamiltonian for Yukawa theory is given by
McCartor and Robertson.\cite{McCartorRobertson}  We
work with a single-fermion truncation\cite{Adelaide} in
which no pair creation or annihilation terms appear and
for which an eigensolution is sought only in the
one-fermion sector.  An analysis of the one-loop fermion
self-energy then shows\cite{ChangYan,PV1} that three
Pauli--Villars bosons are necessary and that their 
couplings to the fermion are fixed as functions of
their masses by three algebraic conditions.  Two of
these bosons must be negatively normed.  A mass
counterterm is also included.
The singularity in the instantaneous fermion contribution
is canceled by addition of an effective 
interaction patterned after the contribution 
of a $Z$ graph.\cite{Adelaide}

Renormalization is done by holding fixed the
mass of the dressed fermion state and by fixing
the value of the expectation value for $:\!\!\phi^2(0)\!\!:$,
where $\phi$ is the boson field operator.  The
eigenvalue problem is rearranged so that the
coefficient of the mass counterterm becomes the
eigenvalue.  In this form the eigenvalue problem
is solved simultaneously with the condition
on $\langle :\!\!\phi^2(0)\!\!:\rangle$ by
iterating in the value of the bare coupling.
This determines the bare mass and bare coupling as
functions of the numerical parameters and the
regularization parameters.  The Fock-state
wave functions are also obtained.

\section{Numerical Methods and Results}
\noindent
The matrix eigenvalue problem is solved with a variant of
the biorthogonal Lanczos method\cite{Lanczos} designed
specifically for an indefinite metric.\cite{Adelaide}
This iterative method requires stopping criteria, which
we take to be convergence of the eigenvalue and of
parts of the boson-fermion wave function, as well.
Because the method generates several eigenvalues,
we also need criteria for selecting the state of
interest.  This state is the ground state, but due
to the indefinite metric, it is not necessarily the
state of lowest mass.  The criteria used for selection
include the following: a positive norm, a real eigenvalue,
absence of nodes in the parallel-helicity boson-fermion
wave function, and a relatively large bare-fermion
probability.  A starting point for the iterations
is generated with use of high-order Brillouin--Wigner
perturbation theory.

An important check on the calculation is found in
the antiparallel boson-fermion wave function, which
due to $J_z$ conservation must be in an $L_z=1$ state.
The calculation does not assume this symmetry but
instead computes the wave function at all $n_x$ and $n_y$
values.  This wave function is found to have the
correct symmetry.

Given a method for the computation of Fock-state
wave functions, any number of interesting
quantities can be subsequently computed.  For example,
matrix elements of the fermion current operator
yield form factors for the dressed state.\cite{FormFactors}

\section{Future Work}
\noindent
The techniques described here are applicable to quantum
electrodynamics and possibly quantum chromodynamics.
In the latter case one would need to use a formulation
such as that of Paston {\em et al}.\cite{Paston} where
an appropriate number of ghost particles and counterterms
are inserted, or perhaps a theory with (broken) supersymmetry.

In the short term, a more complete investigation of Yukawa
theory can be made.  The two-fermion sector of the no-pair
version is of interest because one can consider true
bound states and scattering states.  The
full theory is also interesting because pair terms
bring additional divergences and renormalization of
the boson mass.  These cases can be pursued with direct
extensions of present methods.

\nonumsection{Acknowledgements}
\noindent
The work reported here was done in collaboration 
with S.J. Brodsky and G. McCartor and was supported
in part by the Department of Energy,
contract DE-FG02-98ER41087, and by grants of computing 
time from the Minnesota Supercomputing Institute.

\nonumsection{References}

\end{document}